\documentclass[twocolumn,english,showpacs]{revtex4}
\usepackage{times}
\usepackage[T1]{fontenc}
\usepackage[latin1]{inputenc}
\usepackage{graphicx}
\usepackage{amssymb}

\makeatletter

\usepackage{amsmath}
\usepackage{graphicx}
\usepackage{amssymb}

\usepackage{graphicx}

\usepackage{babel}

\usepackage{babel}
\makeatother
\begin{document}

\title{Thermodynamics of quantum degenerate gases in optical lattices }

\author{P. B. Blakie$^{1}$, A.-M. Rey$^{2}$, and A. Bezett$^{1}$}

\affiliation{$^{1}$Jack Dodd Centre for Photonics and Ultra-Cold Atoms, Department
of Physics, University of Otago, Dunedin, New Zealand}

\affiliation{$^{2}$Institute for Theoretical Atomic, Molecular and Optical Physics,
Harvard-Smithsonian Center of Astrophysics, Cambridge, MA, 02138}

\date{\today{}}

\begin{abstract}
The entropy-temperature curves are calculated for non-interacting
Bose and Fermi gases in a 3D optical lattice. These curves facilitate
understanding of how adiabatic changes in the lattice depth affect
the temperature, and we demonstrate regimes where the atomic sample
can be significantly heated or cooled by the loading process. We assess
the effects of interactions on a Bose gas in a deep optical lattice,
and show that interactions ultimately limit the extent of cooling
that can occur during lattice loading.
\end{abstract}

\pacs{32.80.Pj, 05.30.-d}

\maketitle

\section{Introduction}

\begin{figure}
\includegraphics[width=3.5in,keepaspectratio]{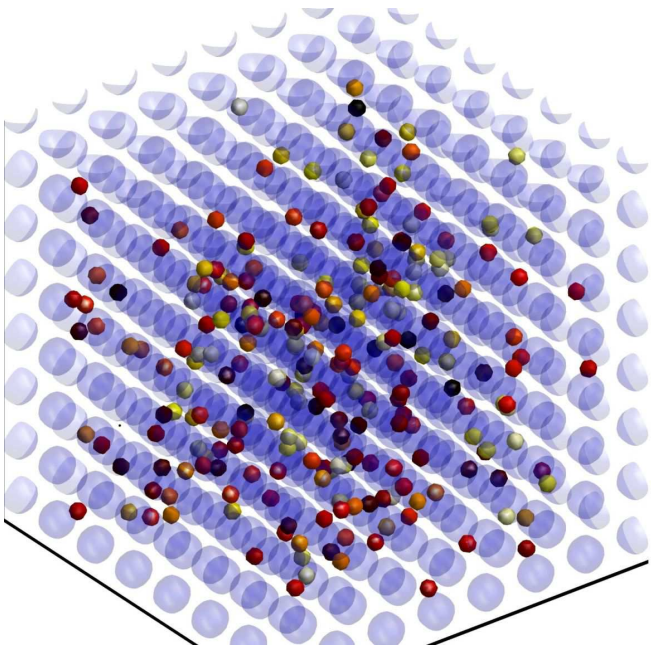}

\caption{\label{fig:LattDiag} Schematic diagram of system under consideration:
a collection of atoms in a uniform lattice of controllable depth. }
\end{figure}

Tremendous advances have been made in the preparation and control
of bosonic and fermionic atoms in optical lattices (e.g. see \cite{Greiner2001a,Greiner2002a,Greiner2002b,Orzel2001a,Anderson1998a,Stoferle2006a,Kohl2005a,Kohl2006a,Schori2004a,Ospelkaus2006a}).
In experiments the gas is typically prepared in the optical lattice
by a slow loading procedure that begins with a weakly trapped gas
and no lattice. During loading, the lattice is turned on in some prescribed
way, and the atoms are localized into the tightly confining potential
wells of the optical lattice. This process is accompanied by a massive
redistribution of the energy states of the system, and it is poorly
understood how the loading process affects the properties of the atoms,
such as their temperature. Many of the physical phenomenon that are
suitable to experimental investigation in optical lattices are sensitive
to temperature and it is therefore of great interest to understand
how the temperature of a quantum degenerate gas changes with lattice
depth. Experimental results by Kastberg \emph{et al.} \cite{Kastberg1995a}
in 1995 showed that loading laser cooled atoms into a three-dimensional
optical lattice caused the atoms to increase their temperature %
\footnote{In fact this study used adiabatic de-loading to reduce the temperature
of the constituent atoms.%
}. Recent studies have shown that there is rich range of behaviour
that can be expected to occur during the loading process at temperatures
much lower than those explored by Kastberg (e.g. see  \cite{Blakie2004a,Blakie2005a,Rey2006a}). 

In this work we compare and contrast the behaviour of Bose and Fermi
systems in optical lattices as a function of the lattice depth. While
some work has been undertaken in the case where an external harmonic
potential is also present (but restricted to a non-interacting tight-binding
approximation) \cite{Kohl2006a}, here we will restrict our attention
to the uniform lattice. Schematically, our system is shown in Fig.
\ref{fig:LattDiag}: a system of atoms confined in a 3D cubic lattice.
As the lattice depth increases the potential changes from that of
a uniform box potential to that of a deep lattice potential. The fundamental
question we wish to address is how the properties of the equilibrium
state, in particular the temperature, change under this loading procedure.
To identify the final thermodynamic state of the system we assume
that the loading procedure is isentropic. In practice experiments
appear to be approximately reversible when the loading is performed
slowly, as has been investigated in Ref. \cite{Gericke2006a}.

The paper is organized as follows: In section \ref{sec:Ideal-Gas-Formalism}
we introduce the theoretical approach we use for calculating the single
particle spectrum and thermodynamic properties of an ideal quantum
gas in an optical lattice. The results of this formalism are presented
and discussed in section \ref{sec:Results}. In section \ref{sec:Effects-of-Interactions}
we address the effects of interactions in application to a bosonic
system, before concluding.

\section{Ideal Gas Formalism\label{sec:Ideal-Gas-Formalism}}

The single particle spectrum completely determines the thermodynamic
properties of an ideal gas. We consider a cubic 3D optical lattice
made from 3 independent sets of counter-propagating laser fields of
wavelength $\lambda$, giving rise to a potential of the form\begin{equation}
V_{{\textrm{Latt}}}(\mathbf{r})=\frac{V}{2}[\cos(2kx)+\cos(2ky)+\cos(2kz)],\label{eq:LattPot}\end{equation}
 where $k=2\pi/\lambda$ is the single photon wavevector, and $V$
is the lattice depth. We take the lattice to be of finite extent with
a total of $N_{s}$ sites, consisting of an equal number of sites
along each of the spatial directions with periodic boundary conditions.
The single particle energies \textbf{$\epsilon_{\mathbf{q}}$} are
determined by solving the Schr\"odinger equation

\begin{equation}
\epsilon_{\mathbf{q}}\psi_{\mathbf{q}}(\mathbf{r})=\frac{\mathbf{p^{2}}}{2m}\psi_{\mathbf{q}}(\mathbf{r})+V_{{\textrm{Latt}}}(\mathbf{r})\psi_{\mathbf{q}}(\mathbf{r}),\label{eq:BlochState}\end{equation}
 for the Bloch states, $\psi_{\mathbf{q}}(\mathbf{r}),$ of the lattice.
For notational simplicity we choose to work in the extended zone scheme
where $\mathbf{q}$ specifies both the quasimomentum and band index
of the state under consideration %
\footnote{For a discussion of how the quantum numbers of quasimomentum and band
index are introduced we refer the reader to Ref. \cite{Mermin1976}%
}. By using the single photon recoil energy, $E_{R}=\hbar^{2}k^{2}/2m$,
as our unit of energy, the energy states of the system are completely
specified by the lattice depth $V$ and the number of lattice sites
$N_{s}$ (i.e. in recoil units $\epsilon_{\mathbf{q}}$ is independent
of $k$).

\begin{figure}
\includegraphics[width=3.3in,keepaspectratio]{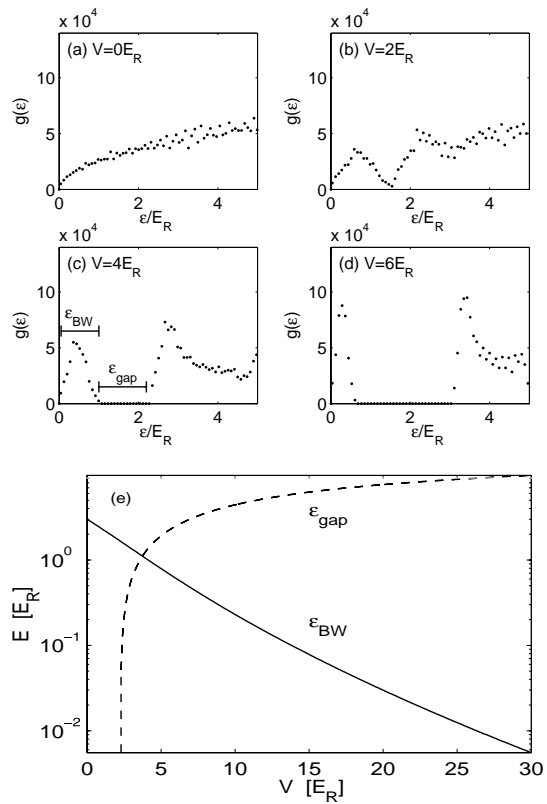}

\caption{\label{cap:Egapbw} (a)-(d) Density of states for a $N_{s}\approx3\times10^{4}$
site cubic lattice at various depths. For a depth of approximately
$V\simeq2E_{R}$ a gap develops in the density of states. In (c) we
illustrate the energy gap $\epsilon_{{\rm gap}}$ and ground band
width $\epsilon_{{\rm BW}}$. Points are determined by numerically
averaging the exact spectrum over a small energy range. (e) The dependence
of the energy gap ($\epsilon_{{\textrm{gap}}}$ dashed line) and ground
band width ($\epsilon_{{\textrm{BW}}}$ solid line) on the lattice
depth (see the text).}
\end{figure}

For completeness we briefly review some important features of the
band structure of Eq. (\ref{eq:BlochState}) relevant to the thermodynamic
properties of the system. The smoothed density of states for the system
for various lattice depths is shown in Figs. \ref{cap:Egapbw}(a)-(d).
For sufficiently deep lattices an energy gap, $\epsilon_{{\textrm{gap}}}$,
will separate the ground and first excited bands (see Fig. \ref{cap:Egapbw}(c)).
For the cubic lattice we consider here, a finite gap appears at a
lattice depth of $V\approx2E_{R}$ %
\footnote{The delay in appearance of the excitation spectrum gap until $V\approx2E_{R}$
is a property of the 3D band structure. In a 1D lattice a gap is present
for all depths $V>0$.%
} (marked by the vertical asymptote of the dashed line in Fig. \ref{cap:Egapbw}(e)).
For lattice depths greater than this, the gap increases with lattice
depth. In forming the gap, higher energy bands are shifted upwards
in energy, and the ground band becomes compressed --- a feature characteristic
of the reduced tunneling between lattice sites. We refer to the energy
range over which the ground band extends as the (ground) band width,
$\epsilon_{{\textrm{BW}}}$ (see Fig. \ref{cap:Egapbw}(c)). As is
apparent in Fig. \ref{cap:Egapbw}(e), the ground band width decreases
exponentially with $V$, causing the ground band to have an extremely
high density of states for deep lattices.

Our primary interest lies in understanding the process of adiabatically
loading a system of $N_{p}$ bosons or fermions into a lattice. Under
the assumption of adiabaticity the entropy remains constant throughout
this process and the most useful information can be obtained from
knowing how the entropy depends on the other parameters of the system.
In the thermodynamic limit, where $N_{s}\rightarrow\infty$ and $N_{p}\rightarrow\infty$
while the filling factor $n\equiv N_{p}/N_{s}$ remains constant,
the entropy per particle is completely specified by the intensive
parameters $T,\, V,$ and $n$. The calculations we present in this
paper are for finite size systems, that are sufficiently large to
approximate the thermodynamic limit. We would like to emphasize the
remarkable feature of optical lattices that $V$ is an adjustable
parameter, in contrast to solid state systems where the lattice parameters
are determined by the constituent atoms and are immutable.

We determine the entropy as follows: We calculate the single particle
spectrum $\{\epsilon_{\mathbf{q}}\}$ for given values of $N_{s}$
and $V$. Working in the grand canonical ensemble we evaluate the
grand partition function $\mathcal{Z}$, according to \begin{equation}
\log\mathcal{Z}=\pm\sum_{\mathbf{q}}\log\left(1\pm e^{-\beta(\epsilon_{\mathbf{q}}-\mu)}\right),\quad\left(_{B}^{F}\right)\label{eq:GrandPot}\end{equation}
 where $\mu$ is found by ensuring particle conservation, and $F$
($B$) refer to the case of fermions (bosons). The entropy of the
system can then be expressed as\begin{equation}
S=k_{B}\left(\log\mathcal{Z}+\beta E-\mu\beta N_{p}\right),\label{eq:Entropy}\end{equation}
 where $\beta=1/k_{B}T$, and $E=-\partial\ln\mathcal{Z}/\partial\beta$
is the mean energy.

Most current Fermi gas experiments are realized with a mixture of
two internal states. This is required because s-wave elastic collisions,
needed for re-equilibration, are prohibited by the Pauli principle
for spin polarized samples %
\footnote{This also means that a single component Fermi gas is quite well described
by a non-interacting theory.%
}. The theory we present here is for the spin polarized case, but is
trivially extensible to multiple components if the lattice potential
is spin independent and the number of atoms in each component is the
same: in this case all extensive parameters are doubled (e.g. $\{ E,S\}$)
and intensive parameters (e.g. $\{ T,\mu\}$) remain the same. The
inclusion of interaction effects, which will be important in the multiple
component case, is beyond the scope of this paper.

\section{Ideal Gas Results\label{sec:Results}}

\subsection{Effect of lattice loading on temperature}

\begin{figure*}
\includegraphics[width=7in,keepaspectratio]{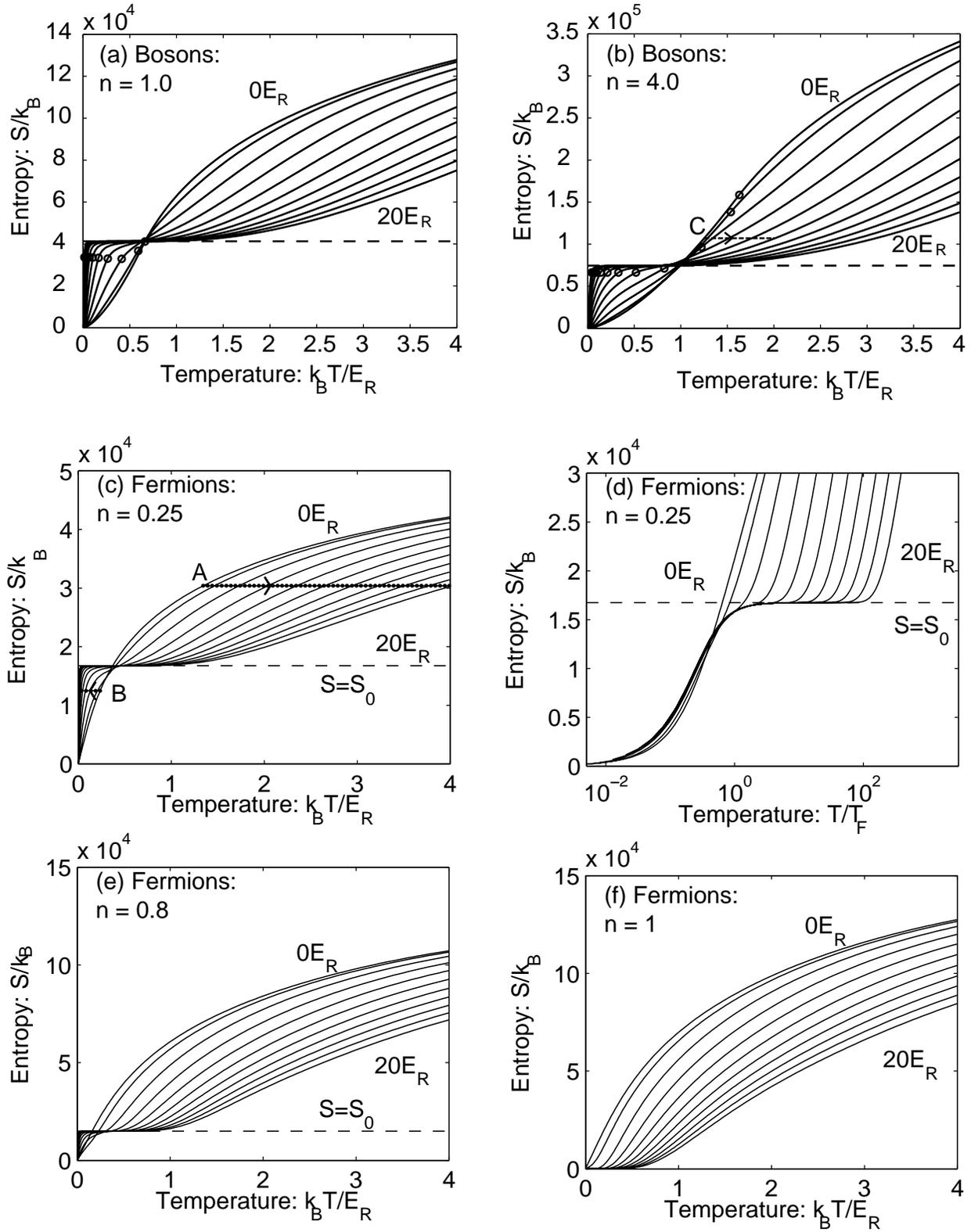}

\caption{\label{FIG:ST1} Entropy versus temperature curves for a $N_{s}\approx3\times10^{4}$
site cubic lattice, at various depths $V=0$ to $20E_{R}$ (with a
spacing of $2E_{R}$ between each curve). Cases considered are (a)
bosons with $n=1.0$, (b) bosons with $n=4.0$, (c) fermions with
$n=0.25,$ (d) fermions (reduced temperature) with $n=0.25$ (e) fermions
with $n=0.8,$ and (f) fermions with $n=1.0$. The entropy plateau
is shown as a dashed line. The processes indicated by the paths labeled
$A,$$B$ and $C$ are discussed in the text. For the bosonic systems
the critical point for condensation on each curve is indicated with
a hollow circle.}
\end{figure*}

In Fig. \ref{FIG:ST1} we show entropy-temperature curves for various
lattice depths and filling factors $n$. These curves have been calculated
for a lattice with $31$ lattice sites along each spatial dimension,
i.e. $N_{s}\approx3\times10^{4}$. A general feature of these curves
is the distinct separation of regions where adiabatic loading causes
the temperature of the sample to increase or decrease, which we will
refer to as the regions of heating and cooling respectively (e.g.
see Figs. \ref{FIG:ST1} (a)-(c) and (e)). These regions are separated
by a value of entropy, $S_{0}$, at which the curves plateau, and
we note that this feature is more prominent on the curves for larger
lattice depths. This plateau entropy is indicated by a horizontal
dashed line and is discussed below. For the case of fermions with
unit filling, shown in Fig. \ref{FIG:ST1}(f), this plateau occurs
at $S_{0}=0$, and only a heating region is observed.

We now explicitly demonstrate the temperature changes that occur during
adiabatic loading using two possible adiabatic processes labeled $A$
and $B$, and marked as dotted lines in Fig. \ref{FIG:ST1}(c). Process
$A$ begins with a gas of free particles in a state with an entropy
value lying above the plateau entropy. As the gas is loaded into the
lattice the process line indicates that the temperature increases
rapidly with the lattice depth. Conversely process $B$ begins with
a gas of free particles in a state with entropy below the plateau.
For this case adiabatic lattice loading causes a rapid decrease in
temperature. This behavior can be qualitatively understood in terms
of the modifications the lattice makes to the energy states of the
system. As is apparent in Fig. \ref{cap:Egapbw}(e), the ground band
rapidly flattens for increasing lattice depth causing the density
of states to be more densely compressed at lower energies. Thus in
the lattice all these states can be occupied at a much lower temperature
than for the free particle case. As we discuss below, for both Bose
and Fermi systems, $S_{0}$ is the maximum entropy available from
only accessing states of the lowest band. If $S<S_{0}$, the temperature
of the system must decrease with increasing lattice depth to remain
at constant entropy. Alternatively, for $S>S_{0}$ the occupation
of states in higher bands is important, and as the lattice depth and
hence $\epsilon_{{\textrm{gap}}}$ increases, the temperature must
increase for these excited states to remain accessible.

\subsubsection{Bosonic systems}

The temperature and entropy at which bosons condense generally changes
with lattice depth, and is indicated by circles on the $S$-$T$ curves
in Figs. \ref{FIG:ST1}(a) and (b). We note that for high filling
factors the condensation points for different lattice depths occur
over a wide entropy range, suggesting that the degree of condensation
will be greatly affected by adiabatic lattice loading. For instance,
consider the adiabatic process indicated by the dashed line and labeled
$C$ in Fig. \ref{FIG:ST1}(b). The system starts as a Bose-condensed
gas of free particles. However, as the lattice depth increases the
condensate fraction decreases until the system passes through the
transition point and becomes uncondensed.

\subsubsection{Fermionic systems}

In addition to the effect that lattice loading has on the absolute
temperature of a Fermi-gas, it is of considerable interest to understand
how the ratio of temperature to the Fermi temperature ($T_{F}$) %
\footnote{The Fermi temperature is given by $T_{F}=\epsilon_{F}/k_{B},$ where
$\epsilon_{F}$ (the Fermi energy) is the energy of the highest occupied
single particle state for the system at $T=0K$.%
} changes. Indeed, the ratio $T/T_{F}$ is the standard figure of merit
used to quantify the degeneracy of dilute Fermi gases. In Fig. \ref{FIG:ST1}(d)
we show how $T/T_{F}$ changes with adiabatic lattice loading for
the same parameters used in Fig. \ref{FIG:ST1}(c). This result indicates
the typical behaviour seen: Below the entropy plateau where cooling
is observed (e.g. see Figs. \ref{FIG:ST1}(c) and (e)), the ratio
of $T/T_{F}$ remains approximately constant, so that there is little
change in the degeneracy of the gas. Above the entropy plateau where
heating was observed, the ratio of $T/T_{F}$ rapidly increases, so
that in this regime the gas will rapidly become non-degenerate as
it is loaded into the lattice. For the unit filled Fermi case (Fig.
\ref{FIG:ST1}(f)), there is no cooling regime, and heating is accompanied
by a rapid increase in $T/T_{F}$ for all initial conditions of the
gas. 

We note that for $n>1$ it is possible to observe a reduction in $T/T_{F}$
for Fermi systems during the loading process. This occurs because
$T_{F}$ increases, because the Fermi energy lies in the excited band.
We do not consider this case here and refer to the reader to Ref.\cite{Blakie2005a}
for details.

\subsection{Entropy plateau\label{sub:Entropy-Plateau}}

In many $S$-$T$ curves a plateau in the entropy is apparent. This
occurs when the gap in the energy spectrum between the states of the
ground and first excited bands is large compared to the energy width
of the ground band, i.e. $\epsilon_{{\rm gap}}>\epsilon_{{\rm BW}}$.
In this case, there is an intermediate temperature range, sufficiently
hot that all the states in the ground band are accessed, yet not hot
enough for states in the next band to be accessed. Within the temperature
range satisfying these conditions, the entropy remains approximately
constant at the value corresponding to the saturated ground band contribution
to the entropy -- we refer to this value as the plateau entropy. We
now provide analytic expressions for this plateau. 

The total number of microstates accessible to the ground band, $\Omega_{0},$
for the cases of fermionic and bosonic particles is given by\begin{eqnarray}
\Omega_{0}^{F} & = & \frac{N_{s}!}{N_{p}!(N_{s}-N_{p})!},\label{eq:OmegaF}\\
\Omega_{0}^{B} & = & \frac{(N_{s}+N_{p}-1)!}{(N_{s}-1)!N_{p}!},\label{eq:OmegaB}\end{eqnarray}
respectively%
\footnote{In order for the Fermi result to hold we require that $N_{p}\leq N_{s}$,
so that the excited band isn't occupied by virtue of the Pauli exclusion
principle.%
}. The corresponding value of entropy $S=k_{B}\log\Omega_{0}$, i.e.
the plateau entropy, is given by \begin{eqnarray}
S_{0}^{F} & \simeq & N_{p}k_{B}\Big[-\frac{1}{n}\log\left(1-n\right)+\log\left(\frac{1}{n}-1\right)\Big],\label{eq:Splateau}\\
S_{0}^{B} & \simeq & N_{p}k_{B}\Big[\log\left(1+\frac{1}{n}\right)+\frac{1}{n}\log\left(1+n\right)\Big],\label{eq:Splateau2}\end{eqnarray}
which we have explicitly written in terms of  the filling factor,
$n\equiv N_{p}/N_{s},$ with the additional validity conditions $1\ll N_{p}\ll N_{s}$
(Fermions) and $N_{s},N_{p}\gg1$ (Bosons). We note an important case
for which the above approximation is invalid is for the Fermi system
with $N_{p}=N_{s},$ i.e. when we have a filling factor of $n=1$,
for which $S_{0}=0$. This case corresponds to the unit filling factor
result shown in Fig. \ref{FIG:ST1}(f) where, as a result of the entropy
plateau occurring at $S_{0}=0,$ only a heating region is observed.

\section{Effects of Interactions in the Bosonic system\label{sec:Effects-of-Interactions}}

We now briefly comment on the role of interactions on the properties
of the Bose gas loaded into an optical lattice. For situations where
the number of bosons is commensurate with the number of lattice sites,
as the lattice depth increases the system will eventually enter the
Mott-insulating state \cite{Jaksch1998a,Greiner2002a}. In this state
the system exhibits a gapped excitation spectrum which is poorly described
by the (gapless) non-interacting spectrum. Here we develop an analytic
description for the unit filled system in the strongly interacting
regime to assess the effects that interactions have on adiabatic loading,
and the behavior of temperature in the deep lattice limit. We also
refer the reader to related results in Ref. \cite{schmidt2006a}. 

In a lattice of depth greater than a few recoils, the system is well-described
by the Bose-Hubbard model \cite{Jaksch1998a}

\begin{equation}
H_{BH}=-J\sum_{\langle\mathbf{i},\mathbf{j}\rangle}\hat{a}_{\mathbf{i}}^{\dagger}\hat{a}_{\mathbf{j}}+\frac{U}{2}\sum_{\mathbf{j}}\hat{a}_{\mathbf{j}}^{\dagger}\hat{a}_{\mathbf{j}}^{\dagger}\hat{a}_{\mathbf{j}}\hat{a}_{\mathbf{j}},\label{eq:HBH}\end{equation}
where $\hat{a}_{\mathbf{j}}$ is the bosonic annihilation operator
of a particle at site $\mathbf{j}=\{ j_{x},j_{y},j_{z}\}$, and the
sum $\langle\mathbf{i},\mathbf{j}\rangle$ is over nearest neighbouring
lattice sites. The interaction parameter $U$ and the tunneling parameter
\textbf{}$J$ can be determined by band structure calculations \cite{Blakie2004b}.

In a deep lattice the tunneling parameter is exponentially suppressed,
and can be taken to be approximately zero %
\footnote{We note that non-negligible $J$ values in the superfluid regime could
be treated using the Hartree-Fock Bogoliubov formalism \cite{Rey2003a,Wild2006a},
however we do not consider this regime here.%
}. In this limit Fock states of the $\hat{a}_{\mathbf{j}}$ operators
diagonalize the Bose-Hubbard Hamiltonian, and in the low temperature
limit $k_{B}T<U$ and for $0<n\lesssim1$ we find that \begin{equation}
\mu=\frac{U}{2}\frac{\left(n-1+\sqrt{3n(2-n)+1}\right)}{2(2-n)}.\label{eq:interMU}\end{equation}
 For the unit filled lattice, $n=1$, the chemical potential is approximately$\mu=U/2$
and one can derive an analytic expression for the grand canonical
partition function given by (see Ref. \cite{Rey2006a} for details) 

\begin{equation}
\mathcal{Z}=\left[1+\frac{e^{\beta\mu}}{2}\left(1+\vartheta_{3}(0,e^{-\beta\mu}\right)\right]^{N_{p}},\label{eq:interZ}\end{equation}
where $\vartheta_{3}(z,q)=1=2\sum_{n=1}^{\infty}q^{n^{2}}\cos(2nz)$
is the Elliptic Theta Function. From these results we can determine
the entropy of the system using Eq. (\ref{eq:Entropy}). 

To assess the effect that interactions have on loading, we use entropy
comparison between the initial system of zero lattice depth (i.e.
the uniform system) and the final deep lattice case described by Eqs.
(\ref{eq:interMU}) and (\ref{eq:interZ}). For the uniform case we
describe the system using the Bogoliubov approximation \cite{Dalfovo1999a},
in which case the quasiparticle spectrum for the uniform system is
given by $\tilde{\epsilon}_{\mathbf{q}}=\sqrt{(\epsilon_{\mathbf{q}}^{0})^{2}+2u\rho\epsilon_{\mathbf{q}}^{0},}$
where $\epsilon_{\mathbf{q}}^{0}=\hbar^{2}\mathbf{q}^{2}/2m$, $\rho=n/(\lambda/2)^{3}$
is the density and $u=4\pi a_{s}\hbar^{2}/m$, with $a_{s}$ the s-wave
scattering length. Assuming that the quasiparticle occupation is given
by the Bose distribution $f(\tilde{\epsilon}_{\mathbf{q}})=[e^{\beta\tilde{\epsilon}_{\mathbf{q}}}-1]^{-1},$
the entropy can be numerically evaluated according to\begin{equation}
S_{{\rm uni}}=k_{B}\sum_{\mathbf{q}}\left[\beta\tilde{\epsilon}_{\mathbf{q}}f(\tilde{\epsilon}_{\mathbf{q}})-\ln\left(1-e^{-\beta\tilde{\epsilon}_{\mathbf{q}}}\right)\right].\label{eq:Suni}\end{equation}

\begin{figure}
\includegraphics[width=3.6in,keepaspectratio]{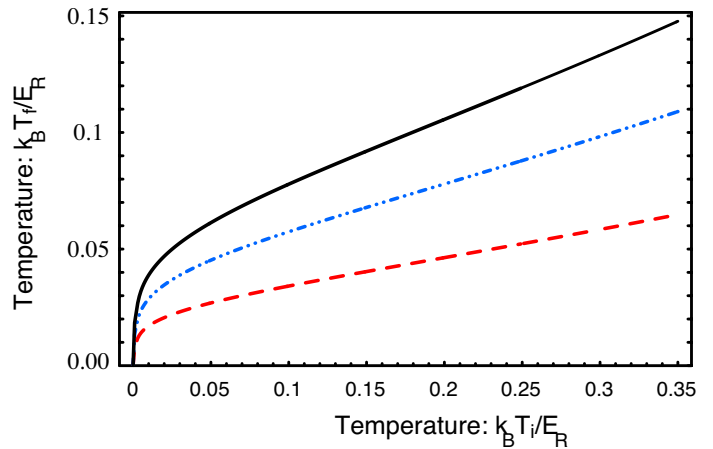}

\caption{\label{fig:Inter} $T_{f}$ versus $T_{i}$ for final lattice depths
of (dashed line) $V=10E_{R}$, (dash-dot line) $V=20E_{R},$ and (solid
line) $V=30E_{R}.$ These results are for the case $n=1$ and for
the parameters of $^{87}$Rb. }
\end{figure}

In Fig. \ref{fig:Inter} we present the results of this entropy comparison,
showing how the initial temperature ($T_{i}$) for the uniform system
relates to the final temperature ($T_{f}$) in the deep lattice under
the assumption of isentropic loading. We notice that $T_{f}$ increases
rapidly as $T_{i}$ changes from zero. This arises because for the
gapless uniform system there is little energy cost to access low energy
modes which contribute the finite initial entropy of the system. In
contrast, in the deep lattice regime, a finite energy cost of order
$U$ must be supplied for the system to access excited states, so
that $k_{B}T\sim U$. Indeed, as is shown in Ref. \cite{Rey2006a},
for $k_{B}T_{i}\gtrsim0.05E_{R},$ and $n=1$, the final temperature
scales linearly with $U$ according to \begin{equation}
T_{f}=\frac{U}{3E_{R}}\left(T_{i}+0.177\frac{E_{R}}{k_{B}}\right).\label{eq:TscaleU}\end{equation}
As $U\sim V^{3/4}$, the temperature must increase with lattice depth,
as is observed in Fig. \ref{fig:Inter}. This is in contrast to the
non-interacting results that show the temperature of the Bose system
(in the low temperature regime where the results of this section hold)
scales as $\epsilon_{{\rm BW}}\sim J$, and as noted earlier this
parameter is exponentially suppressed with increasing lattice depth.
Thus we find that the system has two competing behaviours: for low
lattice depths, the system will be well-described by the non-interacting
result and the temperature will decrease rapidly during loading, as
observed in Figs. \ref{FIG:ST1} (a) and (b). For deeper lattices,
where tunneling between sites is small, the temperature is dominated
by the energy gap of the excitation spectrum, i.e. $U$, which increases
with increasing $V$. Including tunneling effects (i.e. finite $J$)
somewhat suppresses the heating observed in Fig. \ref{fig:Inter}.
This can be qualitatively understood in terms of the modifications
that hopping makes to the eigenstate energies of the system. A non-zero
value of $J$ breaks the degeneracy of the energy gap $U$, leading
to a quasi-band whose width is proportional to $J$. As $J$ increases
(i.e when the lattice becomes shallower) the energy of the lowest
excited states decrease accordingly, while the ground state is only
shifted by an amount proportional to $J^{2}/U$. The lowest energy
excitations then lie closer to the ground state and become accessible
at lower temperatures. As a consequence, the entropy increases (and
thus $T_{f}$ decreases) with respect to the $J=0$ case. Of course
in the deep lattice case, finite $J$ corrections will become vanishingly
small.

\section{Conclusion}

In this paper we have surveyed the physics of loading ultra-cold bosonic
and fermionic atoms into optical lattices. Under the assumption that
this loading is approximately adiabatic (isentropic) we have seen
that there are regimes where the temperature of the system might be
raised or lowered by the loading process. For bosons, the loading
process can be used to reversibly condense the sample. For fermions,
the Fermi energy sets a new energy scale, and for the case of a filled
band no cooling regimes are available. We have examined the effects
of interactions on the Bose system, and seen that in the deep lattice
limit the temperature of the system is proportional to the on-site
interaction strength.

\section*{Acknowledgments}

The authors would like to thank C.W. Clark, J.V Porto and G. Pupillo
for helpful discussions during this research. PBB would like to acknowledge
support from the the Marsden Fund of New Zealand. \bibliographystyle{apsrev}
\bibliography{bandstructbib}

\begin{thebibliography}{22}
\expandafter\ifx\csname natexlab\endcsname\relax\def\natexlab#1{#1}\fi
\expandafter\ifx\csname bibnamefont\endcsname\relax
  \def\bibnamefont#1{#1}\fi
\expandafter\ifx\csname bibfnamefont\endcsname\relax
  \def\bibfnamefont#1{#1}\fi
\expandafter\ifx\csname citenamefont\endcsname\relax
  \def\citenamefont#1{#1}\fi
\expandafter\ifx\csname url\endcsname\relax
  \def\url#1{\texttt{#1}}\fi
\expandafter\ifx\csname urlprefix\endcsname\relax\def\urlprefix{URL }\fi
\providecommand{\bibinfo}[2]{#2}
\providecommand{\eprint}[2][]{\url{#2}}

\bibitem[{\citenamefont{Greiner et~al.}(2001)\citenamefont{Greiner, Bloch,
  Mandel, H{\"a}nsch, and Esslinger}}]{Greiner2001a}
\bibinfo{author}{\bibfnamefont{M.}~\bibnamefont{Greiner}},
  \bibinfo{author}{\bibfnamefont{I.}~\bibnamefont{Bloch}},
  \bibinfo{author}{\bibfnamefont{O.}~\bibnamefont{Mandel}},
  \bibinfo{author}{\bibfnamefont{T.~W.} \bibnamefont{H{\"a}nsch}},
  \bibnamefont{and}
  \bibinfo{author}{\bibfnamefont{T.}~\bibnamefont{Esslinger}},
  \bibinfo{journal}{Phys. Rev. Lett.} \textbf{\bibinfo{volume}{87}},
  \bibinfo{pages}{160405} (\bibinfo{year}{2001}).

\bibitem[{\citenamefont{Greiner
  et~al.}(2002{\natexlab{a}})\citenamefont{Greiner, Mandel, Esslinger,
  H{\"a}nsch, and Bloch}}]{Greiner2002a}
\bibinfo{author}{\bibfnamefont{M.}~\bibnamefont{Greiner}},
  \bibinfo{author}{\bibfnamefont{O.}~\bibnamefont{Mandel}},
  \bibinfo{author}{\bibfnamefont{T.}~\bibnamefont{Esslinger}},
  \bibinfo{author}{\bibfnamefont{T.~W.} \bibnamefont{H{\"a}nsch}},
  \bibnamefont{and} \bibinfo{author}{\bibfnamefont{I.}~\bibnamefont{Bloch}},
  \bibinfo{journal}{Nature} \textbf{\bibinfo{volume}{415}}, \bibinfo{pages}{39}
  (\bibinfo{year}{2002}{\natexlab{a}}).

\bibitem[{\citenamefont{Greiner
  et~al.}(2002{\natexlab{b}})\citenamefont{Greiner, Mandel, H{\"a}nsch, and
  Bloch}}]{Greiner2002b}
\bibinfo{author}{\bibfnamefont{M.}~\bibnamefont{Greiner}},
  \bibinfo{author}{\bibfnamefont{O.}~\bibnamefont{Mandel}},
  \bibinfo{author}{\bibfnamefont{T.~W.} \bibnamefont{H{\"a}nsch}},
  \bibnamefont{and} \bibinfo{author}{\bibfnamefont{I.}~\bibnamefont{Bloch}},
  \bibinfo{journal}{Nature} \textbf{\bibinfo{volume}{419}}, \bibinfo{pages}{51}
  (\bibinfo{year}{2002}{\natexlab{b}}).

\bibitem[{\citenamefont{Orzel et~al.}(2001)\citenamefont{Orzel, Tuchman,
  Fenselau, Yasuda, and Kasevich}}]{Orzel2001a}
\bibinfo{author}{\bibfnamefont{C.}~\bibnamefont{Orzel}},
  \bibinfo{author}{\bibfnamefont{A.~K.} \bibnamefont{Tuchman}},
  \bibinfo{author}{\bibfnamefont{M.~L.} \bibnamefont{Fenselau}},
  \bibinfo{author}{\bibfnamefont{M.}~\bibnamefont{Yasuda}}, \bibnamefont{and}
  \bibinfo{author}{\bibfnamefont{M.~A.} \bibnamefont{Kasevich}},
  \bibinfo{journal}{Science} \textbf{\bibinfo{volume}{23}},
  \bibinfo{pages}{2386} (\bibinfo{year}{2001}).

\bibitem[{\citenamefont{Anderson and Kasevich}(1998)}]{Anderson1998a}
\bibinfo{author}{\bibfnamefont{B.}~\bibnamefont{Anderson}} \bibnamefont{and}
  \bibinfo{author}{\bibfnamefont{M.}~\bibnamefont{Kasevich}},
  \bibinfo{journal}{Science} \textbf{\bibinfo{volume}{282}},
  \bibinfo{pages}{1686} (\bibinfo{year}{1998}).

\bibitem[{\citenamefont{St{\"o}ferle et~al.}(2006)\citenamefont{St{\"o}ferle,
  Moritz, G{\"u}nter, K{\"o}hl, and Esslinger}}]{Stoferle2006a}
\bibinfo{author}{\bibfnamefont{T.}~\bibnamefont{St{\"o}ferle}},
  \bibinfo{author}{\bibfnamefont{H.}~\bibnamefont{Moritz}},
  \bibinfo{author}{\bibfnamefont{K.}~\bibnamefont{G{\"u}nter}},
  \bibinfo{author}{\bibfnamefont{M.}~\bibnamefont{K{\"o}hl}}, \bibnamefont{and}
  \bibinfo{author}{\bibfnamefont{T.}~\bibnamefont{Esslinger}},
  \bibinfo{journal}{Phys. Rev. Lett.} \textbf{\bibinfo{volume}{96}},
  \bibinfo{pages}{030401} (\bibinfo{year}{2006}).

\bibitem[{\citenamefont{K{\"o}hl et~al.}(2005)\citenamefont{K{\"o}hl, Moritz,
  St{\"o}ferle, G{\"u}nter, and Esslinger}}]{Kohl2005a}
\bibinfo{author}{\bibfnamefont{M.}~\bibnamefont{K{\"o}hl}},
  \bibinfo{author}{\bibfnamefont{H.}~\bibnamefont{Moritz}},
  \bibinfo{author}{\bibfnamefont{T.}~\bibnamefont{St{\"o}ferle}},
  \bibinfo{author}{\bibfnamefont{K.}~\bibnamefont{G{\"u}nter}},
  \bibnamefont{and}
  \bibinfo{author}{\bibfnamefont{T.}~\bibnamefont{Esslinger}},
  \bibinfo{journal}{Phys. Rev. Lett.} \textbf{\bibinfo{volume}{94}},
  \bibinfo{pages}{080403} (\bibinfo{year}{2005}).

\bibitem[{\citenamefont{K{\"o}hl}(2006)}]{Kohl2006a}
\bibinfo{author}{\bibfnamefont{M.}~\bibnamefont{K{\"o}hl}},
  \bibinfo{journal}{Phys. Rev. A} \textbf{\bibinfo{volume}{73}},
  \bibinfo{pages}{031601(R)} (\bibinfo{year}{2006}).

\bibitem[{\citenamefont{Schori et~al.}(2004)\citenamefont{Schori, St{\"o}ferle,
  Moritz, K{\"o}hl, and Esslinger}}]{Schori2004a}
\bibinfo{author}{\bibfnamefont{C.}~\bibnamefont{Schori}},
  \bibinfo{author}{\bibfnamefont{T.}~\bibnamefont{St{\"o}ferle}},
  \bibinfo{author}{\bibfnamefont{H.}~\bibnamefont{Moritz}},
  \bibinfo{author}{\bibfnamefont{M.}~\bibnamefont{K{\"o}hl}}, \bibnamefont{and}
  \bibinfo{author}{\bibfnamefont{T.}~\bibnamefont{Esslinger}},
  \bibinfo{journal}{Phys. Rev. Lett.} \textbf{\bibinfo{volume}{93}},
  \bibinfo{pages}{240402} (\bibinfo{year}{2004}).

\bibitem[{\citenamefont{Ospelkaus et~al.}(2006)\citenamefont{Ospelkaus,
  Ospelkaus, Wille, Succo, Ernst, Sengstock, and Bongs}}]{Ospelkaus2006a}
\bibinfo{author}{\bibfnamefont{S.}~\bibnamefont{Ospelkaus}},
  \bibinfo{author}{\bibfnamefont{C.}~\bibnamefont{Ospelkaus}},
  \bibinfo{author}{\bibfnamefont{O.}~\bibnamefont{Wille}},
  \bibinfo{author}{\bibfnamefont{M.}~\bibnamefont{Succo}},
  \bibinfo{author}{\bibfnamefont{P.}~\bibnamefont{Ernst}},
  \bibinfo{author}{\bibfnamefont{K.}~\bibnamefont{Sengstock}},
  \bibnamefont{and} \bibinfo{author}{\bibfnamefont{K.}~\bibnamefont{Bongs}},
  \bibinfo{journal}{Phys. Rev. Lett.} \textbf{\bibinfo{volume}{96}},
  \bibinfo{pages}{180403} (\bibinfo{year}{2006}).

\bibitem[{\citenamefont{Kastberg et~al.}(1995)\citenamefont{Kastberg, Phillips,
  Rolston, Spreeuw, and Jessen}}]{Kastberg1995a}
\bibinfo{author}{\bibfnamefont{A.}~\bibnamefont{Kastberg}},
  \bibinfo{author}{\bibfnamefont{W.~D.} \bibnamefont{Phillips}},
  \bibinfo{author}{\bibfnamefont{S.~L.} \bibnamefont{Rolston}},
  \bibinfo{author}{\bibfnamefont{R.~J.~C.} \bibnamefont{Spreeuw}},
  \bibnamefont{and} \bibinfo{author}{\bibfnamefont{P.~S.}
  \bibnamefont{Jessen}}, \bibinfo{journal}{Phys. Rev. Lett.}
  \textbf{\bibinfo{volume}{74}}, \bibinfo{pages}{1542} (\bibinfo{year}{1995}).

\bibitem[{\citenamefont{Blakie and Porto}(2004)}]{Blakie2004a}
\bibinfo{author}{\bibfnamefont{P.~B.} \bibnamefont{Blakie}} \bibnamefont{and}
  \bibinfo{author}{\bibfnamefont{J.~V.} \bibnamefont{Porto}},
  \bibinfo{journal}{Phys. Rev. A} \textbf{\bibinfo{volume}{69}},
  \bibinfo{pages}{013603} (\bibinfo{year}{2004}).

\bibitem[{\citenamefont{Blakie and Bezett}(2005)}]{Blakie2005a}
\bibinfo{author}{\bibfnamefont{P.~B.} \bibnamefont{Blakie}} \bibnamefont{and}
  \bibinfo{author}{\bibfnamefont{A.}~\bibnamefont{Bezett}},
  \bibinfo{journal}{Phys. Rev. A} \textbf{\bibinfo{volume}{71}},
  \bibinfo{pages}{033616} (\bibinfo{year}{2005}).

\bibitem[{\citenamefont{Rey et~al.}(2006)\citenamefont{Rey, Pupillo, and
  Porto}}]{Rey2006a}
\bibinfo{author}{\bibfnamefont{A.-M.} \bibnamefont{Rey}},
  \bibinfo{author}{\bibfnamefont{G.}~\bibnamefont{Pupillo}}, \bibnamefont{and}
  \bibinfo{author}{\bibfnamefont{J.~V.} \bibnamefont{Porto}},
  \bibinfo{journal}{Phys. Rev. A} \textbf{\bibinfo{volume}{73}},
  \bibinfo{pages}{023608} (\bibinfo{year}{2006}).

\bibitem[{\citenamefont{Gericke et~al.}()\citenamefont{Gericke, Gerbier,
  Widera, Foelling, Mandel, and Bloch}}]{Gericke2006a}
\bibinfo{author}{\bibfnamefont{T.}~\bibnamefont{Gericke}},
  \bibinfo{author}{\bibfnamefont{F.}~\bibnamefont{Gerbier}},
  \bibinfo{author}{\bibfnamefont{A.}~\bibnamefont{Widera}},
  \bibinfo{author}{\bibfnamefont{S.}~\bibnamefont{Foelling}},
  \bibinfo{author}{\bibfnamefont{O.}~\bibnamefont{Mandel}}, \bibnamefont{and}
  \bibinfo{author}{\bibfnamefont{I.}~\bibnamefont{Bloch}},
  \emph{\bibinfo{title}{Adiabatic loading of a bose-einstein condensate in a 3d
  optical lattice}}, \bibinfo{note}{cond-mat/0603590}.

\bibitem[{\citenamefont{Jaksch et~al.}(1998)\citenamefont{Jaksch, Bruder,
  Cirac, Gardiner, and Zoller}}]{Jaksch1998a}
\bibinfo{author}{\bibfnamefont{D.}~\bibnamefont{Jaksch}},
  \bibinfo{author}{\bibfnamefont{C.}~\bibnamefont{Bruder}},
  \bibinfo{author}{\bibfnamefont{J.~I.} \bibnamefont{Cirac}},
  \bibinfo{author}{\bibfnamefont{C.}~\bibnamefont{Gardiner}}, \bibnamefont{and}
  \bibinfo{author}{\bibfnamefont{P.}~\bibnamefont{Zoller}},
  \bibinfo{journal}{Phys. Rev. Lett.} \textbf{\bibinfo{volume}{81}},
  \bibinfo{pages}{3108} (\bibinfo{year}{1998}).

\bibitem[{\citenamefont{Schmidt et~al.}(2006)\citenamefont{Schmidt, Reischl,
  and Uhrig}}]{schmidt2006a}
\bibinfo{author}{\bibfnamefont{K.}~\bibnamefont{Schmidt}},
  \bibinfo{author}{\bibfnamefont{A.}~\bibnamefont{Reischl}}, \bibnamefont{and}
  \bibinfo{author}{\bibfnamefont{G.}~\bibnamefont{Uhrig}},
  \bibinfo{journal}{Eur. Phys. J. D} \textbf{\bibinfo{volume}{38}},
  \bibinfo{pages}{343} (\bibinfo{year}{2006}).

\bibitem[{\citenamefont{Blakie and Clark}(2004)}]{Blakie2004b}
\bibinfo{author}{\bibfnamefont{P.~B.} \bibnamefont{Blakie}} \bibnamefont{and}
  \bibinfo{author}{\bibfnamefont{C.~W.} \bibnamefont{Clark}},
  \bibinfo{journal}{J. Phys. B} \textbf{\bibinfo{volume}{37}},
  \bibinfo{pages}{1391} (\bibinfo{year}{2004}).

\bibitem[{\citenamefont{Dalfovo et~al.}(1999)\citenamefont{Dalfovo, Giorgini,
  Pitaevskii, and Stringari}}]{Dalfovo1999a}
\bibinfo{author}{\bibfnamefont{F.}~\bibnamefont{Dalfovo}},
  \bibinfo{author}{\bibfnamefont{S.}~\bibnamefont{Giorgini}},
  \bibinfo{author}{\bibfnamefont{L.~P.} \bibnamefont{Pitaevskii}},
  \bibnamefont{and}
  \bibinfo{author}{\bibfnamefont{S.}~\bibnamefont{Stringari}},
  \bibinfo{journal}{Rev. Mod. Phys.} \textbf{\bibinfo{volume}{71}},
  \bibinfo{pages}{463} (\bibinfo{year}{1999}).

\bibitem[{\citenamefont{Ashcroft and Mermin}(1976)}]{Mermin1976}
\bibinfo{author}{\bibfnamefont{N.~W.} \bibnamefont{Ashcroft}} \bibnamefont{and}
  \bibinfo{author}{\bibfnamefont{N.~D.} \bibnamefont{Mermin}},
  \emph{\bibinfo{title}{Solid State Physics}} (\bibinfo{publisher}{W.B.
  Saunders Company}, \bibinfo{year}{1976}).

\bibitem[{\citenamefont{Rey et~al.}(2003)\citenamefont{Rey, Burnett, Roth,
  Edwards, Williams, and Clark}}]{Rey2003a}
\bibinfo{author}{\bibfnamefont{A.~M.} \bibnamefont{Rey}},
  \bibinfo{author}{\bibfnamefont{K.}~\bibnamefont{Burnett}},
  \bibinfo{author}{\bibfnamefont{R.}~\bibnamefont{Roth}},
  \bibinfo{author}{\bibfnamefont{M.}~\bibnamefont{Edwards}},
  \bibinfo{author}{\bibfnamefont{C.~J.} \bibnamefont{Williams}},
  \bibnamefont{and} \bibinfo{author}{\bibfnamefont{C.~W.} \bibnamefont{Clark}},
  \bibinfo{journal}{J. Phys. B} \textbf{\bibinfo{volume}{36}},
  \bibinfo{pages}{825} (\bibinfo{year}{2003}).

\bibitem[{\citenamefont{Wild et~al.}(2006)\citenamefont{Wild, Blakie, and
  Hutchinson}}]{Wild2006a}
\bibinfo{author}{\bibfnamefont{B.~G.} \bibnamefont{Wild}},
  \bibinfo{author}{\bibfnamefont{P.~B.} \bibnamefont{Blakie}},
  \bibnamefont{and} \bibinfo{author}{\bibfnamefont{D.~A.~W.}
  \bibnamefont{Hutchinson}}, \bibinfo{journal}{Phys. Rev. A}
  \textbf{\bibinfo{volume}{73}}, \bibinfo{pages}{023604}
  (\bibinfo{year}{2006}).

\end{thebibliography}

\end{document}